\documentclass{article}

\PassOptionsToPackage{authoryear,round,compress}{natbib}
\usepackage[final]{neurips_2021}

\usepackage[utf8]{inputenc} 
\usepackage[T1]{fontenc}    
\usepackage[bookmarks=false,linkcolor=blue,urlcolor=blue,colorlinks,citecolor=blue]{hyperref}       
\usepackage{url}            
\usepackage{booktabs}       
\usepackage{amsfonts}       
\usepackage{mathtools}
\usepackage{nicefrac}       
\usepackage{microtype}      
\usepackage{xcolor}         
\usepackage{xspace}
\usepackage{wrapfig}
\usepackage{todonotes}
\usepackage{caption}
\usepackage{subcaption}

\newcommand{\abs}[1]{\left\lvert#1 \right\rvert}
\newcommand{\norm}[1]{\left\|#1 \right\|}
\newcommand{\eg}{\emph{e.g.}\xspace}
\newcommand{\ie}{\emph{i.e.}\xspace}

\newcommand{\etc}{\emph{etc.}\xspace}

\newcommand{\x}{{\mathbf x}}
\newcommand{\y}{{\mathbf y}}
\newcommand{\z}{{\mathbf z}}

\newcommand{\s}{{\mathbf s}}
\newcommand{\C}{{\mathbb C}}
\DeclareMathOperator*{\EE}{{\mathbb E}}
\DeclareMathOperator*{\argmax}{arg\,max}

\DeclareMathOperator*{\softmax}{Softmax}
\DeclareMathOperator*{\onehot}{OneHot}

\newcommand{\R}{{\mathbb R}}

\title{Towards Neural Variational Monte Carlo \\That Scales Linearly with System Size}

\author{%
  Or Sharir \\
  Caltech\\
  \texttt{ors@caltech.edu} \\
  \And
  Garnet Kin-Lic Chan \\
  Caltech\\
  \texttt{gkc1000@gmail.com} \\
  \And
  Anima Anandkumar \\
  Caltech\\
  \texttt{anima@caltech.edu} \\
}

\begin{document}

\maketitle

\begin{abstract}
    Quantum many-body problems are some of the most challenging problems in science and are central to demystifying some exotic quantum phenomena, e.g., high-temperature superconductors.
    The combination of neural networks (NN) for representing quantum states, coupled with the Variational Monte Carlo (VMC) algorithm, has been shown to be a promising method for solving such problems.
    However, the run-time of this approach scales quadratically with the number of simulated particles, constraining the practically usable NN to — in machine learning terms — minuscule sizes (<10M parameters).
    Considering the many breakthroughs brought by extreme NN in the +1B parameters scale to other domains, lifting this constraint could significantly expand the set of quantum systems we can accurately simulate on classical computers, both in size and complexity.
    We propose a NN architecture called Vector-Quantized Neural Quantum States (VQ-NQS) that utilizes vector-quantization techniques to leverage redundancies in the local-energy calculations of the VMC algorithm~--~the source of the quadratic scaling.
    In our preliminary experiments, we demonstrate VQ-NQS ability to reproduce the ground state of the 2D Heisenberg model across various system sizes, while reporting a significant reduction of about ${\times}10$ in the number of FLOPs in the local-energy calculation.
\end{abstract}

\section{Introduction}

\begin{figure}
    \centering
    \includegraphics[width=\linewidth,keepaspectratio]{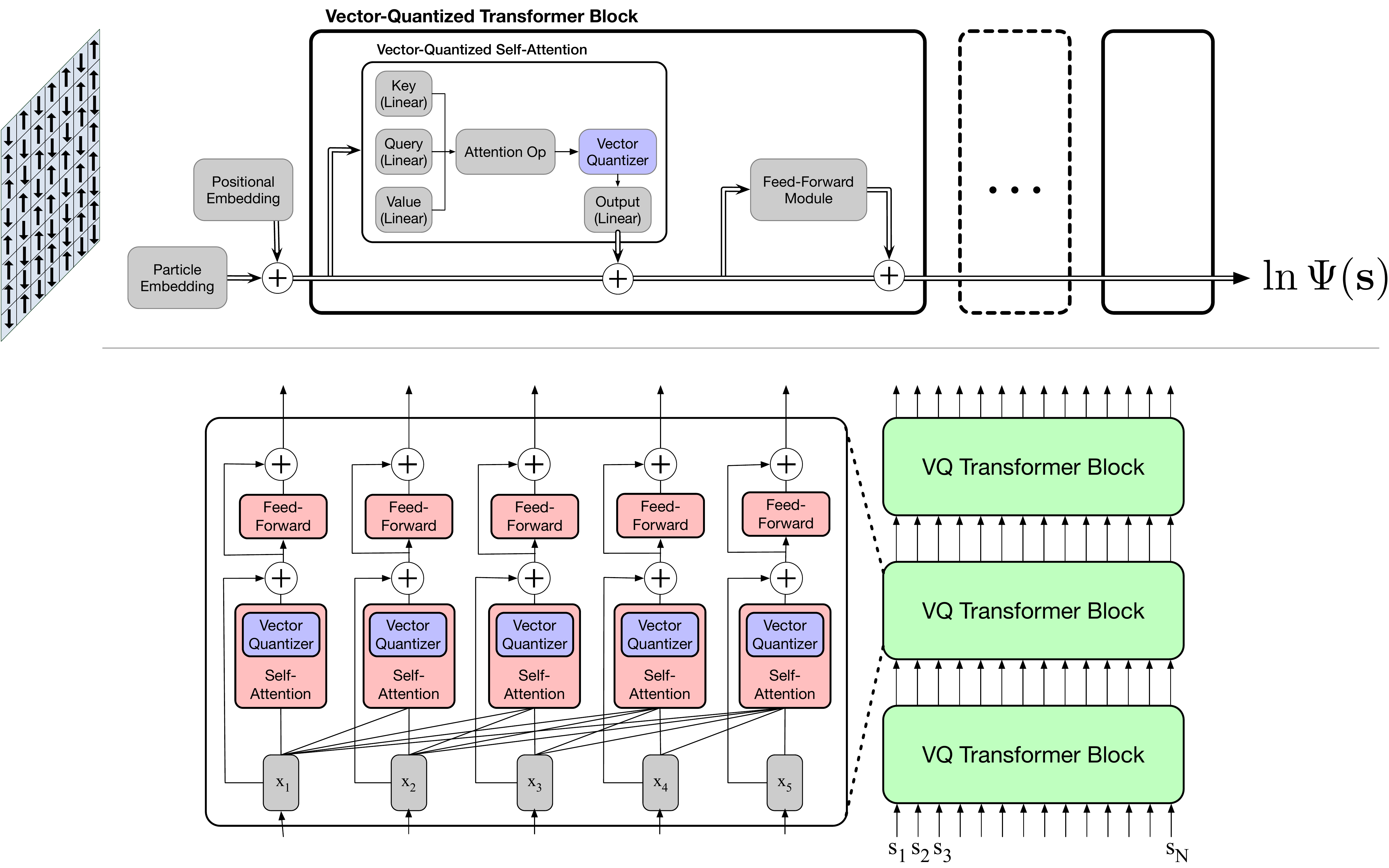}
    \caption{\label{fig:vq-nqs}
        Illustration of the Vector-Quantized Neural Quantum States architecture.
        The top view describes the abstract composition of the network, starting with a lattice of spin particles that are embedded as a sequence of vectors, followed by a sequence of transformer blocks, where we highlight the addition of the vector quantizer following the multi-head self-attention module.
        The bottom view provides the perspective of the transformer architecture as one composed of two main kinds of operations, per-location operations and location-mixer operations.
        The role of the vector quantizers is to help avoid per-location operations when their calculation is redundant, tailor to the VMC algorithm.
    }
\end{figure}

Since its early days, a major challenge in quantum mechanics has been to theoretically understand and model interacting many-body quantum matter.
Many problems in condensed matter, chemistry, nuclear matter, and more are rooted in the intrinsic difficulty of fully representing and manipulating many-body wave functions, which in the worst case grows exponentially with the number of particles.
Over the years, various numerical approaches have been developed.
For example, Tensor Network-based methods~\citep{white1992density,schollwock2011density,orus_tensor_2019,verstraete2008matrix,cirac_matrix_2020} rely on a specific family of ansatz amenable to high-ordered linear manipulations, giving rise to very efficient optimization algorithms but at the tradeoff of a more limited expressivity~\citep{nqs-vs-tns}.
On the other hand, Variational Monte Carlo (VMC)~~\citep{mcmillan_ground_1965} could in principle be used with any parameterized wave-function ansatz, but different function families could result in different expressivity-efficiency tradeoffs.

Recently, neural-network-based techniques have been proposed~\citep{carleo2017solving} as an alternative ansatz for the VMC algorithm, dubbed Neural Quantum States (NQS), aiming to tap into their remarkable expressive power~\citep{nqs-vs-tns,levine2019quantum,deng2017quantum,PhysRevB.97.085104}.
This approach has been demonstrated to be quite competitive for describing many systems in both physics~\citep{PhysRevX.8.011006,choo_two-dimensional_2019,sharir_deep_2020,schmitt_quantum_2020,hibat-allah_recurrent_2020,torlai_neural-network_2018} and chemistry~\citep{pfau_ab_2020,hermann_deep-neural-network_2020,choo_fermionic_2020}.
However, solving many-body problems with VMC\&NQS comes with an inherent computational cost that grows quadratically\footnote{When using some of the latest techniques, e.g., \citet{sharir_deep_2020,hibat-allah_recurrent_2020}.
} with the system size (number of bodies / particles), denoted henceforth by $N$.
These costs end up limiting the practical uses of NQS to small sizes by machine learning (ML) terms.
While recent breakthroughs in ML rely on NN in the range of +1B parameters~\citep{dalle2,GPT3}, and can go as high as 1T parameters~\citep{SwitchTransformers}, the NN used for representing quantum states are limited to just millions of parameters~\citep{sharir_deep_2020}~--~without resorting to very large clusters that are inaccessible to most.
The expressiveness afforded by large NN could be the key factor in successfully modeling the ground states of some elusive quantum systems.

In this paper, we propose a novel NQS architecture that aims to mitigate this limitation.
It is built upon the Transformers architecture~\citep{transformers}, but with \emph{vector quantizers}~\citep{VQVAE} appended to the output of the self-attention modules, and so we call it Vector-Quantized NQS, or VQ-NQS for short.
The main feature of this setup is its ability to detect redundancies in its computational graph, and thus avoid recomputing the same operations.
By quantizing intermediate representations, we can represent them in a compressed form of indices pointing to unique vectors, akin to sparse-matrix formats.
This allows us to apply most NN operations just on the unique set of vectors, rather than on the entire hidden state.
This feature is especially important for VMC, as its most taxing step is calculating \emph{local energies}, which involve evaluating a NN over a large set of nearly identical inputs.
VQ-NQS can exploit this redundancy in the inputs to compute the local energies with significantly less resources, which under some assumptions reduces the overall cost of VMC to nearly linear.

We present preliminary experimental results using VQ-NQS, and comparing it to the conventional NQS ansatz.
Since VQ-NQS is still a work-in-progress, we focus our experiments on two key aspects: (i) demonstrate this ansatz is sufficiently expressive to represent the ground states of well-studied systems, and (ii)~demonstrate the significantly reduced computational cost of calculating local energies under these VQ-NQS ground-state representations.
We use the conventional VMC\&NQS to find close approximations to the ground state of the antiferromagnetic two-dimensional Heisenberg model for various system sizes, followed by NN distillation techniques to find a similar VQ-NQS representation.
Our results conclusively demonstrate that VQ-NQS can recover the ground state to very high precision, obtaining relative error on the order of $10^{-4}$ or less, while capable of about ${\times}10$ reduction in the computational cost (measured in FLOPs) of computing local energies.
We leave for followup works the integration and evaluation of VQ-NQS method under the full VMC algorithm.
Nevertheless, our preliminary results already demonstrate the significant potential of VMC\&VQ-NQS~--~taking one step closer towards making neural-based VMC scale linearly with the system size.
This would open the door to large-scale NQS, simulating larger systems, and bringing us closer to solving some of the yet intractable Hamiltonians.

The rest of the paper is organized as follows.
In sec.~\ref{sec:backgound} we provide the necessary background on Variational Monte Carlo, Neural Quantum States, and the computational bottleneck located in the local-energy calculations.
In sec.~\ref{sec:vq-nqs} we introduce our VQ-NQS architecture, and describe how it can be compute local energies more efficiently by leveraging redundancies in its computational graph.
We present our preliminary experimental results in sec.~\ref{sec:exp}, and conclude in sec.~\ref{sec:summary} where we discuss our main findings and future directions.

\section{Variational Monte Carlo, Neural Quantum States, and the Local Energy Bottleneck}\label{sec:backgound}

We consider in the following a pure quantum system, constituted by $N$ discrete degrees of freedom (\eg spins, occupation numbers, \etc) ${\s{\equiv}(s_1,\ldots,s_N)}$, where $s_i \in \{q_1, \ldots,q_d\}$, such that the wave-function~(WF) complex-valued amplitudes $\Psi(\s)$ fully specify its state.
$\Psi:d^N\to\C$ can be viewed both as a function and as a high-dimensional vector with multi-index coordinates denoted by $\s$.
Given the Hamiltonian matrix $H \in \C^{d^N \times d^N}$ corresponding to some model of a quantum system, we define the \emph{ground state} of $H$ as the wave-function corresponding to its minimal eigenvector $H\Psi_0 = E_0 \Psi_0$, where $E_0$ is the ground-state energy.
If not for the high dimensional ($d^N$) nature of the problem, finding the ground state would have been trivial.

The Variational Monte Carlo algorithm reframes the problem of finding the ground state as a stochastic optimization problem.
We start with the Rayleigh Quotient method, i.e., $\Psi_0 = \arg\min_{\Psi} \frac{\Psi^\dagger H \Psi}{\Psi^\dagger \Psi}$, where for a generic trial wave function $\Psi$, we define $E(H, \Psi) = \frac{\Psi^\dagger H \Psi}{\Psi^\dagger \Psi}$ as its energy.
Given some parameterized function $\Psi_\theta$, we can rewrite its energy as an expectation over the probability $P(\s) = \nicefrac{\Psi(\s)^2}{\Psi^\dagger \Psi}$ (known as the Born probability):
\begin{align}
    E(H, \Psi) & = \EE_{\s}\left[E_{loc}(\s;H,\Psi)\right], \label{eq:vmc} \\ E_{loc}(\s;H,\Psi) & \equiv \sum_{s'} H_{\s,\s'} \frac{\Psi(\s')}{\Psi(\s)}, \label{eq:local-energy}
\end{align}
where $E_{loc}(s;H,\Psi)$ is known as the local energy of the configuration $s$.
Henceforth we would omit $H$ and $\Psi$ from $E$ and $E_{loc}$ where their context is clear.
While the local energy term might seem intractable, because most Hamiltonians of interest are very sparse with roughly $O(N)$ non-zero entries per row (e.g., for local Hamiltonians) then calculating $E_{loc}(s)$ requires merely evaluating $\Psi$ on just $O(N)$ coordinates.
With the expectation in eq.~\ref{eq:vmc}, we can then estimate the energy of a given state by drawing samples according to its corresponding Born probability without requiring access to all its coordinates.
Finally, we can estimate the energy gradient with the log-derivative trick and plug it in to any stochastic gradient descent optimizer to find the ground state of $H$.
The exact form for estimating gradient over $K$ samples is given by:
\begin{align}
    \frac{\partial E}{\partial \theta} & \approx \sum_{i=1}^K (E_{loc}(\s^{(i)}) - E)^*\frac{\partial \log \Psi_\theta(\s^{(i)})}{\partial \theta}
\end{align}

Following the approach introduced in \citet{carleo2017solving}, we can represent the logarithm of the trial wave function $\log(\Psi_\theta(\s))$ by some neural network function $f_\theta:d^N \to \C$.
If we have an efficient method for sampling according to $f$ then can plug it into the VMC algorithm mentioned above.
In the general case, we would have to resort to a Metropolis-Hasting type of sampling algorithm, which could require a polynomial number of evaluations for each sample.
Instead, we follow \citet{sharir_deep_2020} for an NQS architecture that supports efficient sampling with the cost of single evaluation.
Namely, we use the polar decomposition, i.e., $\Psi(\s)=\sqrt{P(\s)}\exp(i \phi(\s))$, to represent the wave function with two separate real functions, a normalized distribution represented by an autoregressive network $P(\s)$ and a phase network $\phi(\s)$.
$P$ represents the distribution via the chain rule $P(\s) = \prod_i P(s_i|s_{<i})$, and so supports direct sampling with the same cost as computing $P(\s)$.

When considering the costs of VMC\&NQS using autoregressive networks, the cost of sampling as well as the cost of computing $\frac{\partial \log \Psi_\theta(\s)}{\partial \theta} $ are proportional to the cost of a single forward pass each, whereas the cost of the computing the local energy term for every sample is equal to $O(N)$ forward passes.
Since each forward pass processes an $N$-length input, then for most commonly used NN architectures (\ie, convolutional, recurrent, transformers) the cost of each forward pass would (roughly) be at least $\Omega(N\cdot \abs{\text{params}})$.
Hence the total cost of a single VMC iteration is at best $O(N^2 \cdot \abs{\text{params}})$, \ie, quadratic in the system size.

As should be apparent, the quadratic scaling is driven by the linear number of forward passes for computing the local energy.
Let us examine the local-energy calculation more carefully for the specific and common case of a $k$-local Hamiltonian, \ie, $H=\sum_{i=1}^m H^{(i)}$ where local term $H^{(i)}$ operates on at most $k$ locations $J^{(i)}$, where if any $s_j \neq s'_j$ for $j\not\in J^{(i)}$ then $H^{(i)}_{\s,\s'} = 0$.
For every row $\s$, each local term contributes at most $O(d^k)$ non-zero entries that are identical to $s$ except for (at most) the $k$ locations of $J^{(i)}$.
In total, we get $O(m d^k)$ non-zero entries per row, where the typical values are $d=2$, $k\in\{1,2,3\}$, and $m = O(N)$.
To give a concrete example, in the case of the Ising spin model, where $s$ is a bit string (\ie, $d=2$), for each row $s$ the non-zero entries are found at $s$ itself or $s$ with one of its bits flipped.

Ideally, a function operating on these very similar sets of inputs should be able to reuse a large portion of its calculations.
Indeed, consider the case of autoregressive networks, where the $P(\s)$ is defined by the chain rule and we already computed all the conditional probabilities for $\s$, and now we wish to compute $P(\s')$ where $s'_j = 1 - s_j$.
For $i < j$ all locations are equal, and so we could reuse the conditional probabilities $P(s_i|s_{<i})$, however, all conditional probabilities for $i\geq j$ would have to be recomputed to take the changed spin into consideration.
On average, we could save half the computational cost by leveraging this redundancy.
Nevertheless, such cost savings still leave us with a linear dependence on $N$.
In the next section, we will describe an architecture for which we can reuse most of the computational graph if only a few input locations are changed, reducing the total cost to sub-linear in $N$.

\section{Vector-Quantized Neural Quantum States}\label{sec:vq-nqs}

In this section we describe our proposed ansatz, VQ-NQS, that leverages the redundancy in the local-energy calculations to decrease the overall runtime to be nearly linear in the system size.

\subsection{Architecture}

VQ-NQS is based on the Transformers architecture~\citep{transformers}, where non-overlapping groups of particles corresponds to the ``tokens'' in the original architecture.
A group of $K$ particles can be represented by a ``vocabulary'' embedding of $d^K$ ``word''-vectors of dimension $d_{\text{hidden}}$.
These groups are similar to the patches used in Vision Transformer~\citep{VisionTransformers}.
After mapping groups of particles to embedding vectors, including their respective positional embeddings, they are processed by a sequence of transformer blocks composed of self-attention modules, which mixes information across token location, and feed-forward modules, which process each location separately.
We direct readers to prior works for a detail description of these common elements.

As discussed in sec.~\ref{sec:backgound}, we represent the wave function $\Psi(\s)$ with two separate networks, a distribution network $P(\s)$ and a phase network $\phi(\s)$.
We use a unidirectional decoder transformer network to represent the distribution as an autoregressive network, using the particle groups as the variables of the distribution.
We use a bidirectional encoder transformer network to represent the phase, where we pool the hidden states of the last layer, followed by a linear projection reducing it to a scalar.

Our architecture deviates from the conventional transformers architecture, by appending a vector quantization~(VQ)~\citep{VQVAE} layer at the end of the self-attention module, just before the linear output projection and the residual connection.
See illustration in fig.~\ref{fig:vq-nqs}.
The VQ layer is applied to each location separately, akin to the feed-forward module, and maps each hidden input vector to the nearest vector in its codebook comprising $Q$ learnable vectors.
Our VQ layer implementation combines many of the recent innovations proposed in the literature~\citep{mama2021nwt,zeghidour2021soundstream,yu2022vectorquantized}, including using a variant of the straight-through Gumbel-Softmax gradient estimator, dead code removal, and multi-head VQ.
Specifically, we use the following function for the forward pass:
\begin{align}
    D_{hj}                & = -\norm{\x_h - W_{h,j,:}}_2,                   \\
    I_{\text{VQ}}(\x;W)_h & = \argmax_j W_{hj},                             \\
    V_{\text{VQ}}(\x;W)_h & = W_{h, I_{\text{VQ}}(\x;W)_h,:}, \label{eq:vq}
\end{align}
where $\x \in \R^{d_{\text{input}}}$ is split into $H$ segments of length $\nicefrac{d_{\text{input}}}{H}$ denoted by $\x_1,\ldots,\x_H$, and $W \in \R^{H \times Q \times \nicefrac{d_{\text{input}}}{H}}$ is a multi-head codebook of $Q$ vectors per each of the $H$ segments.
$V_{\text{VQ}}(\x;W)$ is the quantized vector returned by the layer, where $I_{\text{VQ}}(\x;W)$ are the indices to the codebook that we will use for bookkeeping purposes in the redundant computation stage.
To learn the codebook $W$, we can rewrite eq.~\ref{eq:vq} as:
\begin{align}
    V_{\text{VQ}}(\x;W)_h & = g(D_h)^T W_h, \\
    g(\z)                 & = \mathrm{SG}\!\left[\onehot\left(\argmax\nolimits_j z_j\right)\right] +  \softmax(\z) -\mathrm{SG}\left[\softmax(\z)\right]
    \label{eq:st}
\end{align}
where $\onehot(j)$ is a vector with one in the $j$'th coordinate and zeros elsewhere, and $\mathrm{SG}\!\left[ \cdot \right]$ is the stop gradient operation, preventing backward computation in the autodiff alogrithm.
Eq.~\ref{eq:st} can be viewed as a form of a straight-through gradient estimator for the discrete forward pass.
In terms of capacity, both $Q$ and $H$ control how much information can be passed through the quantization bottleneck, which can be summarized as the total number of bits $H\cdot \log_2(Q)$.
Notice that varying $H$ does not affect the runtime of the forward pass, while it does increase the information bandwidth of the VQ layer.

\subsection{Avoiding Redundant Calculations}

As in sec.~\ref{sec:backgound}, let us consider the case of a set $\s^{(1)},\ldots,\s^{(K)} \in d^N$ of similar inputs, such that $K = O(N)$ and there exists $M \ll N$ so that for any $i\neq j$ it holds that $1 \leq \abs{\{t: s^{(i)}_t \neq s^{(j)}_t\}} \leq M$, and we wish to compute the values of $\Psi(\s^{(i)})$.
Ideally, the similarity in the inputs should be reflected in the computational graph of $\Psi$, resulting in redundant calculations that could be avoided.
We will shortly demonstrate how the VQ-NQS architecture facilitate exactly that.

To simplify the discussion, let us consider VQ-NQS where each particle, i.e., a specific coordinate $s^{(k)}_i$, correspond to a single ``token''.
Each token is then mapped to a vector $\x^{(k)}_i$ according to the ``word'' and positional embeddings, which will then be processed by the transformer blocks (whether they be of the encoder or decoder type) to result in $N$ output vectors $\y^{(k)}_i$.
We will focus on reducing the complexity of this vector to vector mapping, as it is shared by both the $P(\s)$ and $\phi(\s)$ networks.

Let us denote with $\x^{(k,l)}_i$ the hidden states of the network after the $l$'th transformer block, where $\x^{(k)}_i = \x^{(k,0)}_i$.
In the case above, each particle is mapped to one of $U \leq M(K - 1) + N$ possible vectors.
Hence, instead of representing the full $K\cdot N$ vectors $\{\x^{(k)}_i\}_{k,i}$, we can represent them in the compressed form as a pair of tensors $(I, V)$, where $I$ are the indices in the $\{0,\ldots,U - 1\}$ range pointing to their corresponding vectors $V$ from the unique set.
Notice that any operation that we wish to apply identically over all locations (e.g., the feed-forward transformer modules, or LayerNorm layers) can be directly applied on this compressed format by simply applying the operation on $V$.
This is correct since all locations that point to the same vector will result in the same output vector after applying the location-identical operation.

Using the above compressed representation, if we had simply removed all of the self-attention modules from the network and only kept the per-location operations, then we could have computed the entire mapping of the $K\cdot N$ input vectors using just $\nicefrac{U}{KN} = O(\nicefrac{M}{N})$ of the total cost, i.e., with the same cost as just $O(M)$ forward passes.
However, since the self-attention module mixes information across locations, its output space is no longer redundant and compressible.
It is here where the VQ modules come into play, compressing the self-attention output space by limiting its information bandwidth.
Quantizing the output space means outputs that where previously nearly equivalent are now exactly so, and it gives us the unique identifiers that allow for efficient conversion to the compressed format.
We should emphasize that it is only for similar inputs that we can reasonably expect the VQ layer to result in a compressible output space, effectively assuming a small Lipschitz constant of the learned mapping prior to quantization.
While this property could be more explicitly enforced with various forms of regularizations, in our experience this was not needed.
Instead, we control the compression of the output space through the bandwidth of the VQ layer (i.e., with $Q$ and $H$), with the aim of keeping the compression at an $O(\nicefrac{M}{N})$ rate.

Though promising, we should stress that the above method only avoids the redundancy with the per-location operations.
While computing the attention matrix itself could have similar runtime benefits by examining only the unique key-query inner-products, multiplication of the attention matrix with the value matrix still requires the full cost of dense multiplication.
This specific limitation could be avoided by employing one of the linear attention mechanisms.
However, for the specific use case of representing NQS this is typically not needed because $N$ is typically in the regime where the cost of the Transformer architecture is dominated by the per-location operations.
As a concrete example, in most cases of interest $N < 500$, and after using groups of 4 the effective sequence length is below 125, so for a hidden size of 256 the attention costs is less than $\nicefrac{1}{13}$ of the total cost~--~in our experiments it is even less than $1\%$.
As for the cost of VQ itself, due to the dense output of the attention module prior to quantization, it is still applied to the entire $KN$ set of vectors, but by using multiple heads ($H>1$) and a small $Q$, we can keep its cost relatively small while still being sufficiently expressive.

In conclusion, we can estimate the cost of evaluating $\Psi$ on this set as nearly linear with $N$.
Though asymptotically the cost will still depend on $K$, by taking the constants into account we can neglect those terms in our regime of interest, and thus arrive to a more realistic cost estimation closer to $O(M)$ forward passes.
Since every forward pass itself mostly depends linearly on $N$ and since $M$ is a small constant typically less than 4, then the total cost of evaluating $\Psi$ on this set is nearly linear with $N$.
As we demonstrate in sec.~\ref{sec:exp}, we can indeed achieve significant savings calculating local energies using VQ-NQS while approximating ground states of well-studied systems with high precision.

\section{Experiments}\label{sec:exp}

\begin{figure}
    \centering
    \begin{subfigure}[b]{0.49\linewidth}
    	\centering
    	\includegraphics[width=\linewidth,keepaspectratio]{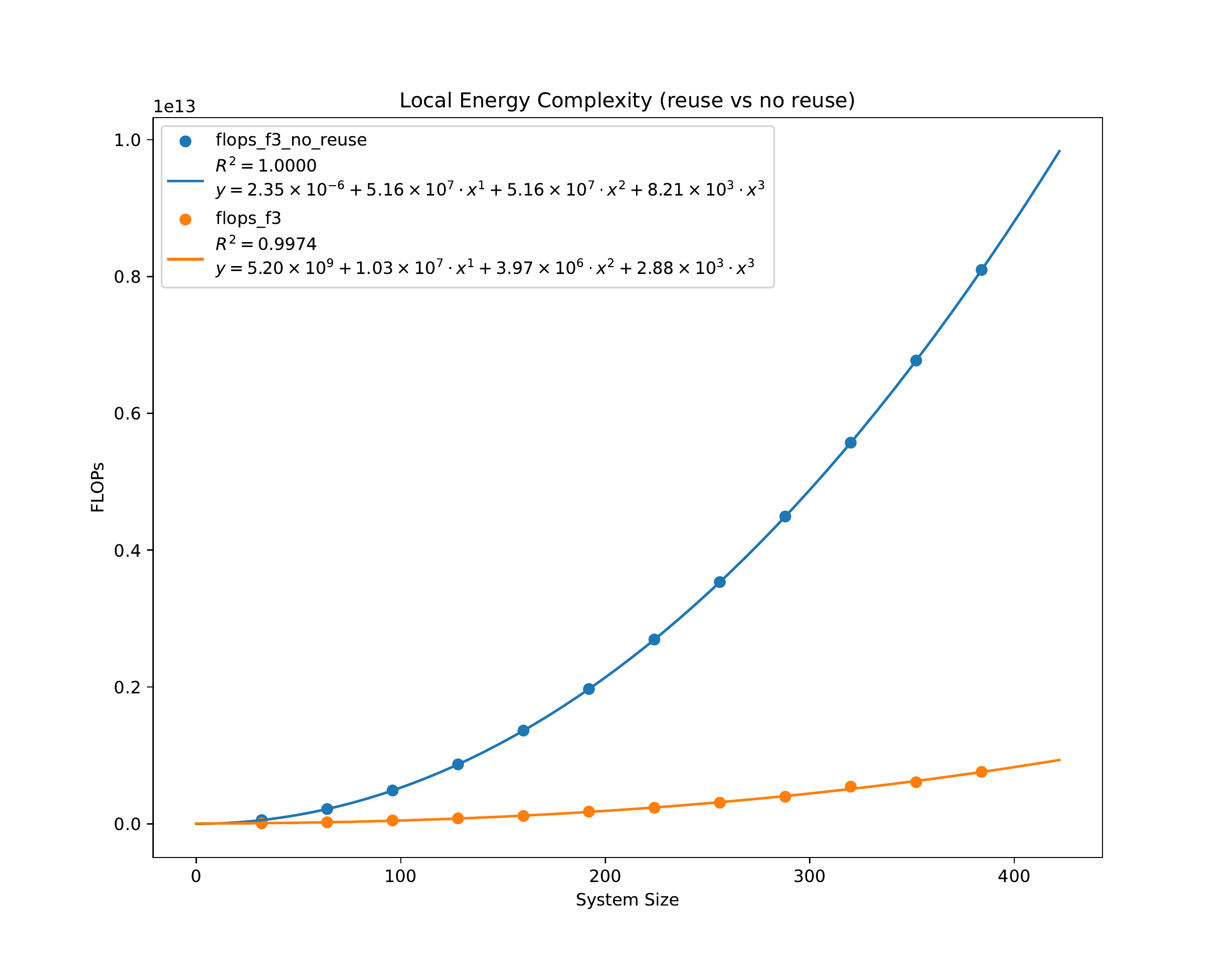}
    	\caption{Reuse vs. no reuse of computation.}
        \label{fig:ising1d:all}
    \end{subfigure}
    \begin{subfigure}[b]{0.49\linewidth}
    	\centering
    	\includegraphics[width=\linewidth,keepaspectratio]{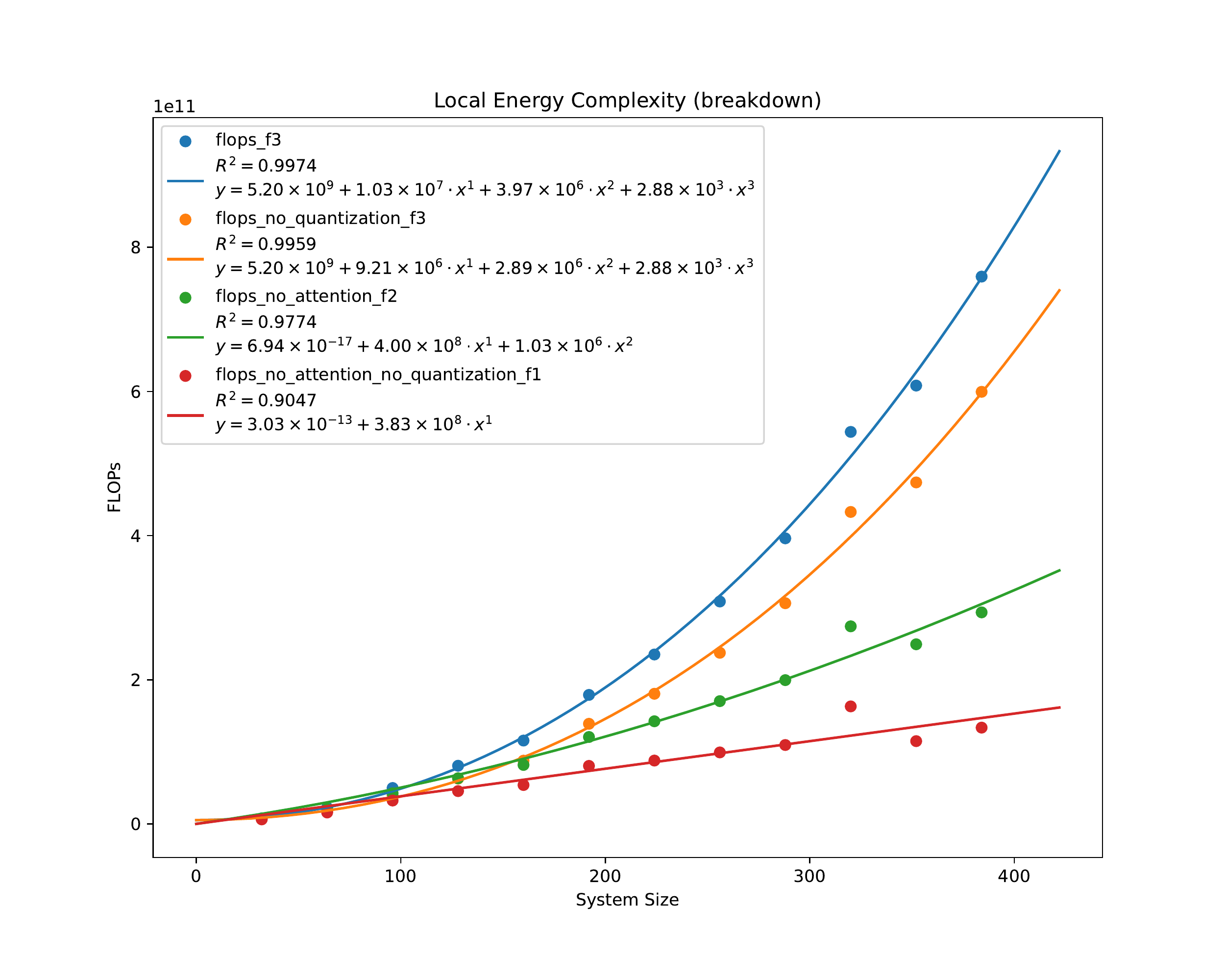}
    	\caption{Breakdown to different components.}
        \label{fig:ising1d:breakdown}
    \end{subfigure}
    \caption{\label{fig:ising1d}
        Demonstrating the scaling of the computational cost on the one-dimensional transverse-field Ising model. We use a VQ-NQS using one transformer block, with a hidden size of 256, 8 self-attention heads, a quantizer with 4 heads and 64 codebook vectors each. Across all system size we achieve at least a $10^{-5}$ relative error of ground state energy. In (a) we present the scaling of FLOPs of the model, comparing the scaling without reusing computation and with reuse. A true linear scaling is not yet expected with our initial implementation. To demonstrate the potential of our approach, in (b) we breakdown the cost (when reusing computation) to show that if we omit the not (yet) optimized parts of the network, then our method does indeed follow a linear scaling. 
    }
\end{figure}

\begin{table}
    \caption{\label{table:results}
        Comparing a baseline transformer-based NQS versus VQ-NQS with various quantization bandwidths on the two-dimensional Heisenberg model.
        Both networks consist of three transformer blocks, plus a half-block composed of just a self-attention module at the end.
        They use a hidden size of 256 and 8 self-attention heads.
        For the VQ-NQS network, we use VQ layers with either 1, 2, 4, or 8 heads, each with a codebook of 64 vectors, thereby amounting to 6 bits, 12 bits, 24 bits, or 48 bits of quantization bandwidth.
        We omit the VQ layer from the last half-block as it does not bring any runtime benefits.
        The ground-state energy relative error is estimated by ground state energies recovered by either exact diagonalization in the $4 \times 4$ case, or Quantum Monte Carlo~\citep{bauer_alps_2011} for the larger lattices, as computed by prior works~\citep{liu2017gradient,PhysRevB.103.235155}.
        The FLOPs are estimated on a batch of 512 samples for which the local energies are calculated in parallel, where the reported FLOPs are averaged over 20 such batches.
        The total savings is simply the ratio between the VQ-NQS networks and the baseline, whereas the quantized ops savings is evaluated by only considering operations that are applied to quantized inputs.
    }
    \centering
\begin{tabular}{rccccc}
\toprule
Model  & Energy & Relative Error & FLOPs & Total Savings & Quantized-Ops Savings  \\
\midrule
\multicolumn{6}{c}{ $4 \times 4$ Lattice }  \\  
\midrule
  Ref. Energy & -0.574325 & - & - & - & - \\
           Baseline & -0.574326 & 1.8e-06 & 2.07e+11 & - & - \\
 6b  VQ-NQS & -0.574309 & 2.8e-05 & 1.49e+10 & $\times$13.9 &  $\times$20.1 \\
12b  VQ-NQS & -0.574325 & 4.8e-07 & 1.18e+10 & $\times$17.5 &  $\times$28.7 \\
24b  VQ-NQS & -0.574325 & 4.0e-07 & 1.69e+10 & $\times$12.3 &  $\times$16.8 \\
48b  VQ-NQS & -0.574326 & 1.8e-06 & 1.68e+10 & $\times$12.3 &  $\times$16.9 \\
\midrule
\multicolumn{6}{c}{ $6 \times 6$ Lattice }  \\  
\midrule
  Ref. Energy & -0.603522 & - & - & - & - \\
           Baseline & -0.603512 & 1.6e-05 & 1.11e+12 & - & - \\
 6b  VQ-NQS & -0.603495 & 4.5e-05 & 9.42e+10 & $\times$11.8 &  $\times$16.9 \\
12b  VQ-NQS & -0.603506 & 2.6e-05 & 1.91e+11 & $\times$5.8 &  $\times$6.8 \\
24b  VQ-NQS & -0.603506 & 2.7e-05 & 2.90e+11 & $\times$3.8 &  $\times$4.2 \\
48b  VQ-NQS & -0.603509 & 2.1e-05 & 3.48e+11 & $\times$3.2 &  $\times$3.5 \\
\midrule
\multicolumn{6}{c}{ $8 \times 8$ Lattice }  \\  
\midrule
  Ref. Energy & -0.619040 & - & - & - & - \\
           Baseline & -0.619003 & 6.0e-05 & 3.63e+12 & - & - \\
 6b  VQ-NQS & -0.618961 & 1.3e-04 & 4.00e+11 & $\times$9.1 &  $\times$12.6 \\
12b  VQ-NQS & -0.618986 & 8.7e-05 & 7.07e+11 & $\times$5.1 &  $\times$6.1 \\
24b  VQ-NQS & -0.618991 & 7.9e-05 & 1.14e+12 & $\times$3.2 &  $\times$3.5 \\
48b  VQ-NQS & -0.618992 & 7.7e-05 & 1.44e+12 & $\times$2.5 &  $\times$2.7 \\
\midrule
\multicolumn{6}{c}{ $10 \times 10$ Lattice }  \\  
\midrule
  Ref. Energy & -0.628667 & - & - & - & - \\
           Baseline & -0.628593 & 1.2e-04 & 9.10e+12 & - & - \\
 6b  VQ-NQS & -0.628505 & 2.6e-04 & 9.81e+11 & $\times$9.3 &  $\times$14.1 \\
12b  VQ-NQS & -0.628562 & 1.7e-04 & 1.75e+12 & $\times$5.2 &  $\times$6.4 \\
24b  VQ-NQS & -0.628578 & 1.4e-04 & 2.65e+12 & $\times$3.4 &  $\times$3.9 \\
48b  VQ-NQS & -0.628592 & 1.2e-04 & 3.50e+12 & $\times$2.6 &  $\times$2.8 \\
\midrule
\multicolumn{6}{c}{ $12 \times 12$ Lattice }  \\  
\midrule
  Ref. Energy & -0.635203 & - & - & - & - \\
           Baseline & -0.635084 & 1.9e-04 & 1.91e+13 & - & - \\
 6b  VQ-NQS & -0.634985 & 3.4e-04 & 2.17e+12 & $\times$8.8 &  $\times$14.5 \\
12b  VQ-NQS & -0.635045 & 2.5e-04 & 3.30e+12 & $\times$5.8 &  $\times$7.7 \\
24b  VQ-NQS & -0.635064 & 2.2e-04 & 4.80e+12 & $\times$4.0 &  $\times$4.8 \\
48b  VQ-NQS & -0.635072 & 2.1e-04 & 6.75e+12 & $\times$2.8 &  $\times$3.2 \\
\bottomrule
\end{tabular}
\end{table}

In this section we report our preliminary experimental results of using VQ-NQS for representing ground states and exploring their potential to reduce the complexity of VMC with neural networks.
To this end, we compare our VQ-NQS architecture using various quantization bandwidths against a baseline transformer-based NQS.
The two architectures are identical except for the addition of vector quantizers in the former.
The two fundamental properties we examine are (i)~the expressive power of VQ-NQS for approximating ground states, and (ii)~the potential savings this architecture could bring for calculating local energies, and thus for VMC as well.

To establish these two properties, we focus on a two well-studied quantum system for which precise solutions can be obtained, namely, the one-dimensional transverse-field Ising model and the two-dimensional antiferromagnetic Heisenberg model, both with open boundary conditions. Specifically, the corresponding Hamiltonians are defined as $H = -J \sum_{<i, j>} \sigma^i_z \sigma^j_z - \Gamma \sum_i \sigma^i_x$ and  $H = \sum_{<i, j>} \sigma^i_x \sigma^j_x + \sigma^i_y \sigma^j_y + \sigma^i_z \sigma^j_z$, respectively, where the $<i, j>$ denote summing over nearest neighbors on the 1/2D lattice and $\sigma^i_x, \sigma^i_y, \sigma^i_z$ are the typical Pauli matrices operating on the $i$'th site.
In both cases we can ensure the solution to be strictly positive (for Heisenberg using the Marshal sign rule), and therefore we only train the spin distribution networks, and set the phase to zero.
Our training procedure is as follows. We begin by using the regular VMC algorithm to optimize our baseline for finding a highly-precise ground state approximation, on the order of $10^{-4}$ or less across all system sizes. We then distill this ground-state approximation to our VQ-NQS model, where we iteratively sample a batch of spin configurations and their corresponding log-amplitudes and then train our VQ-NQS model to match these values by minimizing the $L_2$ loss objective. Finally, we can measure the cost (in FLOPs) of computing the local energy of our VQ-NQS model.

Given its simplicity, the 1D Ising model is ideal for demonstrating the scaling of our proposed model to very large system sizes, up to 384 spins. We focus on the case of $\Gamma = 1.0$ at the critical point of the model. Across all system sizes our model converges to at least $10^{-5}$ relative error of the ground state energy. We present the FLOPs of VQ-NQS with and without taking advantage of the redundancy afforded by its quantized feature maps to reuse computation. We accompany our plots with their corresponding best-fitted polynomials to highlight their empirical scaling. Our results and hyperparameters are shown in figure~\ref{fig:ising1d:all}. Given that VQ-NQS uses a quadratic attention module over a relatively small sequence length, then we expect (and observe) the case without reuse to follow a cubic scaling pattern with a small coefficient for the cubic monomial. When reusing computation, we still observe a cubic scaling, but the coefficients for non-linear terms are much smaller, resulting in a significant overall reduction in FLOPs. As mentioned in section~\ref{sec:vq-nqs}, our initial implementation does not yet take full advantage of the quantized feature maps to improve the cost of the entire network. Specifically, the self-attention operation is not yet optimized (leading to the cubic term), which as result also affect the cost of quantization itself (leading to the quadratic term). We highlight this in figure~\ref{fig:ising1d:breakdown} that provides a breakdown of the FLOPs by subtracting different elements from the architecture. This shows that when considering only the optimized parts of the network, we do observe a clear linear scaling. 

Next, we demonstrate that our model is able to cope with a more intricate Hamiltonian, namely, the two-dimensional Heisenberg model. We test our model on lattices size between $4 \times 4$ and $12 \times 12$. Both the baseline and VQ-NQS achieve at least $10^-4$ relative error.
While better precision is possible by training for longer and using extensive hyper-parameter search~\citep{sharir_deep_2020}, these results are sufficient for our preliminary needs.
We report our results in table~\ref{table:results}, which also include the hyper-parameters of our networks.
According to our results, even the most constrained VQ-NQS networks with a quantization bandwidth of just 6 bits can capture the ground states and achieve nearly the same precision as our unconstrained NQS baselines.
Moreover, VQ-NQS can achieve theoretical savings of almost $\times$10 across all lattices, proving its potential to speedup VMC iterations.
Notice how by changing the quantization bandwidth we can control the trade-off between the ansatz expressivity and its computational efficiency.
As with the Ising model, we wish to draw attention to the last column of the table that only takes into account the optimized parts of our network.
In the case of the completely tractable $4 \times $ lattice where the networks are clearly overspecified, the VQ layers effectively learn to skip over most of the operations as is shown by the up to $\times$29 savings.
These results indicate that more performance could be extracted by infusing the architecture with additional VQ layers, or by changing the self-attention mechanism to other ``token mixers''~\citep{MLPMixer,MetaFormer,FLASH} that might better lend themselves to incorporate VQ elements.

\section{Discussion}\label{sec:summary}

In this work we laid the groundwork toward a neural VMC algorithm that scales linearly with the system size.
The bottleneck of the algorithm in prior methods is located in the local-energy calculations, which involve the evaluation of the wave-function ansatz on a large set of similar inputs.
Our approach is to leverage this redundancy in the input space by a neural network that combines the Transformers architecture and vector quantization layers to avoid needless computation of redundant neurons, which we call VQ-NQS.
As our complexity analysis predicts and as can be observed from our initial experimental results, large savings can be attained using this approach.

Nevertheless, we acknowledge that the method described in this paper has some limitations that prevent it from reaching its full potential and obtaining true linear scaling.
The main drawback is that not all operations in the network can take advantage of the redundancy afforded by the vector quantization elements.
When we focus on just the operations that apply to quantized inputs, the gains are much more substantial.
Adapting the architecture to accommodate that is the central path to achieve linear scaling, without the caveats articulated in sec.~\ref{sec:vq-nqs}.
Furthermore, while our experiments demonstrated the essential properties and the promise of using VQ-NQS, a complete evaluation of utilizing it within the VMC framework is needed to establish a reduction in the total cost of reaching some target precision of the ground state energy, and for a diverse set of quantum systems.
Eliminating these limitations and further evaluation of this method will be left to a future follow-up work.

As a final note, we would like to emphasize that while VQ-NQS was designed to solve quantum many-body problems, the fundamental concepts at its core could be applied to many other domains.
Namely, a lot of problems have a large redundancy in their input space that can potentially be exploited, e.g., processing video frames or editing text documents, which have natural spatiotemporal redundancies.
Extending the VQ-NQS architecture to such domains could provide a fruitful avenue for future research.

%

\bibliographystyle{plainnat}
\bibliography{refs}


\end{document}